# Kinetics of charge carrier recombination in $\beta$-Ga$_2$O$_3$ crystals

T. T. Huynh[1], L. L. C. Lem[1], A. Kuramata[2], M. R. Phillips[1] and C. Ton-That[1,*]

[1] School of Mathematical and Physical Sciences, University of Technology Sydney, Ultimo, NSW 2007, Australia

[2] Novel Crystal Technology, Inc., Sayama, Saitama 350-1328, Japan

* Corresponding author. Email: cuong.ton-that@uts.edu.au

**Abstract**

Cathodoluminescence spectra were measured to determine the characteristics of luminescence bands and carrier dynamics in $\beta$-Ga$_2$O$_3$ bulk single crystals. The CL emission was found to be dominated by a broad UV emission peaked at 3.40 eV, which exhibits strong quenching with increasing temperature; however, its spectral shape and energy position remain virtually unchanged. We observed a super-linear increase of CL intensity with excitation density; this kinetics of carrier recombination can be explained in terms of carrier trapping and charge transfer at Fe impurity centres. The temperature-dependent properties of this UV band are consistent with weakly bound electrons in self-trapped excitons with an activation energy of 48 ± 10 meV. In addition to the self-trapped exciton emission, a blue luminescence (BL) band is shown to be related to a donor-like defect, which increases significantly in concentration after hydrogen plasma annealing. The point defect responsible for the BL, likely an oxygen vacancy, is strongly coupled to the lattice exhibiting a Huang-Rhys factor of ~ 7.3.

# I. INTRODUCTION

Monoclinic β-Ga$_2$O$_3$ has recently gained increasing attention as an emerging wide bandgap semiconductor with a high breakdown electric field (~ 8 MV/cm) and attractive electronic and optical properties for use as the active material in power electronic and optoelectronic devices.[1,2] Most β-Ga$_2$O$_3$ crystals and epilayers available today exhibit auto n-type conductivity, with the ionization energy of dominant donors being estimated to be ~ 40 meV,[3] and has widely been attributed to oxygen vacancies ($V_o$).[4] However, this assignment is inconsistent with theoretical studies, which predict $V_o$ to be relatively deep at ~ 1 eV below the conduction band minimum (CBM).[5,6] An alternative explanation for the n-type conductivity is from the inadvertent incorporation of impurities, especially Sn or H, that act as shallow donors.[5] Additionally, theoretical and experimental studies have suggested that gallium vacancies ($V_{Ga}$) are native acceptors and are responsible for the compensation of n-type conductivity.[7-9] Recently, substitutional Fe at Ga sites (Fe$_{Ga}$) has been found to be an energetically favourable acceptor defect in edge-defined film-fed grown (EFG) β-Ga$_2$O$_3$ crystals, which dominates deep-level defect traps.[10] It is well known that Fe acts as a highly efficient capture centre for both electrons and holes in group-II and -III compound semiconductors;[11] however, thus far there is little information available about the influence of Fe impurities on carrier kinetics and compensation in β-Ga$_2$O$_3$. In addition, controlling the charge state of Fe centres in β-Ga$_2$O$_3$ may be interesting for spintronics because of the recent prediction of room temperature ferromagnetism.[12]

As with all new wide bandgap semiconductors, a detailed knowledge of the optical and electronic structure is crucial for the design and optimization of Ga$_2$O$_3$-based devices. Information on carrier kinetics is important for Ga$_2$O$_3$ applications in photovoltaics and photodetectors as photo-response time is strongly influenced by carrier trapping. With an intrinsic band gap of ~ 4.9 eV, a broad UV-blue optical emission in Ga$_2$O$_3$ centered ~ 3.3 eV widely reported in the literature with a large Stokes shift of ~ 1.6 eV.[13,14] For highest quality β-Ga$_2$O$_3$ single crystals without luminescent transitions associated with impurities, the optical emission features UV and blue luminescence (BL) bands with energies in the range of 3.1-3.6 and 2.5-2.8 eV respectively, whose intensity ratio depends on excitation conditions and temperature.[14,15] The UV band has been shown to be independent of dopants or sample preparation methods and accordingly attributed to intrinsic origin. Computational and Electron Paramagnetic Resonance (EPR) studies clearly demonstrated that self-trapped holes are thermally stable in β-Ga$_2$O$_3$,[16,17] which serve as a precursor for the formation of self-trapped excitons (STEs) that are responsible for the dominant UV emission in β-Ga$_2$O$_3$ single crystals.[3,15] The BL emission has been found to be strong in conducting Ga$_2$O$_3$ samples and as a result has been assigned to the recombination of an electron bound to a $V_O$ defect,[3,9,14] Several Ga$_2$O$_3$-based devices have been reported with an optical response in the UV range that exhibit a dramatic change in the electronic transport with increasing operation temperature,[2,18] indicating the importance of controlling and understanding temperature-dependent carrier recombination kinetics in Ga$_2$O$_3$. In this work we investigate characteristics of luminescence bands in Ga$_2$O$_3$ and provide evidence of the involvement of carrier trapping in an electron injection-induced optical emission.

## II. EXPERIMENTAL DETAILS

Experiments were conducted on EFG β-Ga$_2$O$_3$ rectangular single crystals fabricated by Novel Crystal Technology Inc., Japan.[19] The crystal has a (-201) surface orientation and dimensions of $10 \times 15 \times 0.68$ mm$^3$. The main impurities in these crystals are Si, Ir, Al and Fe, with the [Fe] in the ppm ($10^{16}$ - $10^{17}$ cm$^{-3}$) range,[10,19] which was independently verified by Inductively Coupled Plasma Mass Spectrometry (ICP-MS) for the sample used in this work. X-ray diffraction analysis confirmed a single crystal monoclinic structure (not shown). Some Ga$_2$O$_3$ samples were incorporated with hydrogen using remote hydrogen radio-frequency plasma (15 W, sample temperature 470 K). The crystal was characterised by scanning cathodoluminescence (CL) spectroscopy using a FEI Quanta 200 scanning electron microscope (SEM) equipped with a parabolic mirror collector and an Ocean Optics QE65000 CCD array spectrometer. For temperature-dependent CL spectroscopy, the crystal was mounted cold and hot stages in the SEM, which enables measurements between 10 and 500 K. All luminescence spectra were corrected for the total system response.

## III. RESULTS AND DISCUSSION

The temperature-resolved CL spectra of the β-Ga$_2$O$_3$ crystal with identical excitation conditions (15 kV, 8 nA), shown in Fig 1(a), reveals a broadening of the lower energy side of the emission due to the enhancement of the BL component with increasing temperature. In contrast, the spectral shape of the normalized UV band attributed to STEs is virtually unchanged as evidenced from the invariance of the high energy side of the UV peak with increasing temperature. The temperature independence of the spectral shape of the UV band is commonly observed for excitons immobilised by a local deformation of crystal lattice because the energy of a STE emission critically depends on the distortion energy of self-trapped holes and is only weakly influenced by the band edge energies.[20,21] The overall CL spectrum of Ga$_2$O$_3$ is slightly red shifted from 3.45 at 10 K to 3.28 eV at 360 K due to the overlap with the enhanced BL emission relative to the UV peak. Over the same temperature range the BL/UV intensity ratio increases from 0.15 to 0.73. It is also important to note that both the UV and BL bands were measurable up to ~ 500 K. The β-Ga$_2$O$_3$ BL has been reported by other workers and attributed to $V_O$ or ($V_O$, $V_{Ga}$) vacancy pair.[3,14] Depth-resolved CL analysis reveals identical peak shape and energy position with increasing acceleration voltage (see the spectra in Appendix A), ruling out any contribution of surface effects to the CL emission. In analogy to other semiconductors exhibiting STE luminescence,[21,22] holes in β-Ga$_2$O$_3$ can be trapped as a small polaron and strongly coupled to the lattice, resulting in a broad, Gaussian-like band as expected due to the involvement of a large number of phonons. Based on these considerations, the Gaussian peak fitting of the Ga$_2$O$_3$ spectra was made with the UV component fitted to the invariant high energy side of the UV peak, giving rise to its peak position of $E_{UV}$ = 3.40 ± 0.05 eV and FWHM$_{UV}$ = 0.70 ± 0.05 eV for all temperatures up to 400 K. Conversely, the BL component was found to broaden and is red shifted from 3.05 eV at 10 K to 2.92 eV at 600 K. To probe the characteristics of the BL luminescent centre, the temperature dependent width of the emission band is fitted according to the expression below derived from the configuration coordinate model. The FWHM of the BL is ~ 0.55 eV at temperatures below 40 K and monotonically increase with temperature [Fig 1(d)]. The Huang-Rhys factor, $S$, and the effective phonon energy, $\hbar\omega$, were obtained from fitting the FWHM to the following equation:[23]

$$FWHM = 2.36 S\hbar\omega (\coth\left(\frac{\hbar\omega}{2kT}\right))^{1/2} \qquad [1]$$

The curve in Fig 2(d) correspond to the best fit parameters to the BL width: $S$ = 7.3 ± 0.7 and $\hbar\omega$ = 32 ± 4 meV. These corresponds to a Frank-Condon shift, $S\hbar\omega$, of 234 meV. This measured phonon energy is within the low range of energies for longitudinal optical (LO) phonons obtained by spectroscopic ellipsometry.[24] However, the curve fit based on the Eq. [1] provides the effective phonon energy, which is an average of all phonons involved in the optical transition.

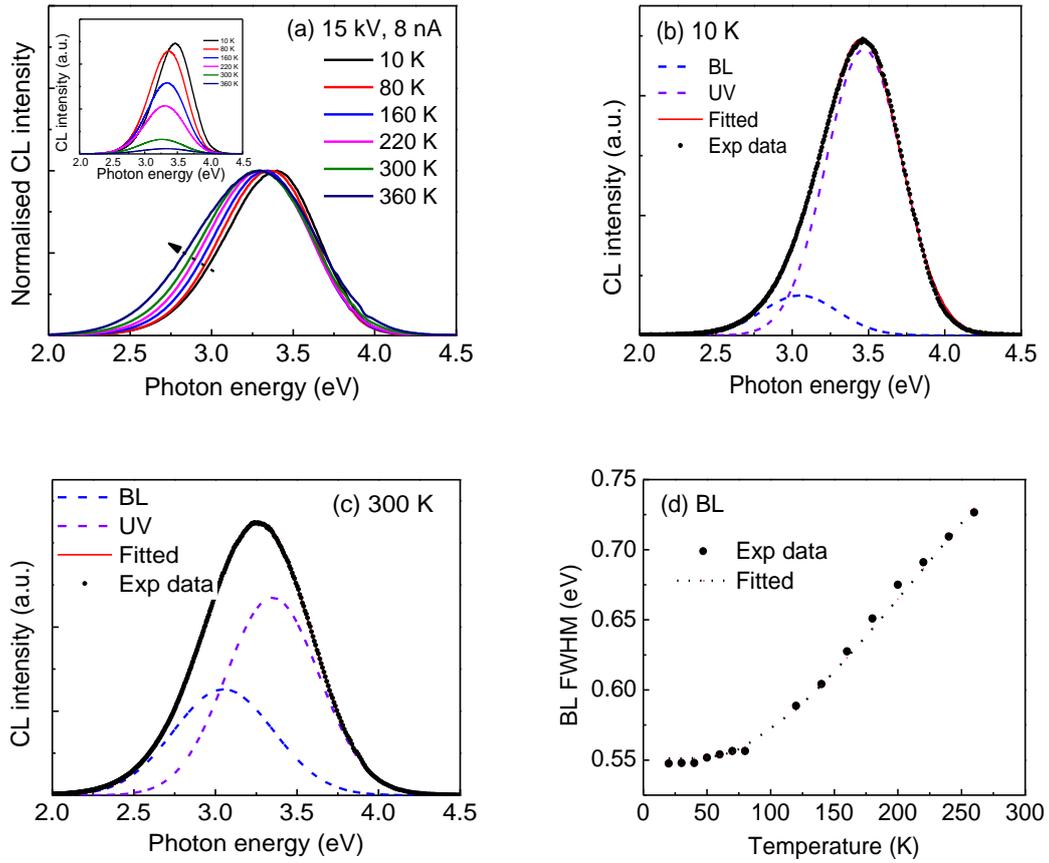

**Figure 1**. (a) Temperature-resolved CL spectra of (-201) β-Ga$_2$O$_3$ crystal under identical excitation conditions (15 kV, 8 nA), showing the broadening of the lower energy side of the spectrum with increasing temperature. Inset: the CL intensity decreases in intensity due to phonon scattering. (b, c) CL spectra fitted with two Gaussians for the UV and BL bands using comparable peak energy and FWHM values in previous reports.[14,25] The UV component can be fitted to the nearly temperature-insensitive high energy side of the CL spectrum with $E_{UV}$ = 3.40 ± 0.05eV, FWHM$_{UV}$ = 0.73 ± 0.05 eV (Δλ = 80 nm). (d) The FWHM of the BL modelled using the configuration coordinate description of phonon coupling to point defects, yielding the mean phonon energy $\hbar\omega$ = 32 ± 4 meV and Huang Rhys factor S = 7.3 ± 0.7, consistent with a typical for defects with strong electron- phonon coupling.

In order to probe the physical nature of defects responsible for the BL in $Ga_2O_3$, the crystal, was doped with $H^+$ from a remote hydrogen plasma at 470 K; a temperature where atomic H is highly diffusive in $Ga_2O_3$.[26] Under the plasma treatment conditions similar to those used in this work, hydrogen has been shown to be incorporated into $Ga_2O_3$ to a depth of > 500 nm.[26] The remote plasma treatment was found to cause no noticeable changes in both the crystal structure (confirmed by X-ray diffraction) and surface morphology (confirmed by AFM), but the BL is significantly enhanced following the H incorporation (Figure 2). The BL/UV intensity ratio increases by an order of magnitude from 0.03 for undoped $Ga_2O_3$ to 0.25 after 40 minutes plasma (inset of Figure 2). This enhancement (i) is consistent with an increase in the concentration of $V_O$ produced by H ions, which can extract surface oxygen atoms[27] and (ii) supports the assignment of the BL to recombination of $V_O$-bound electrons.

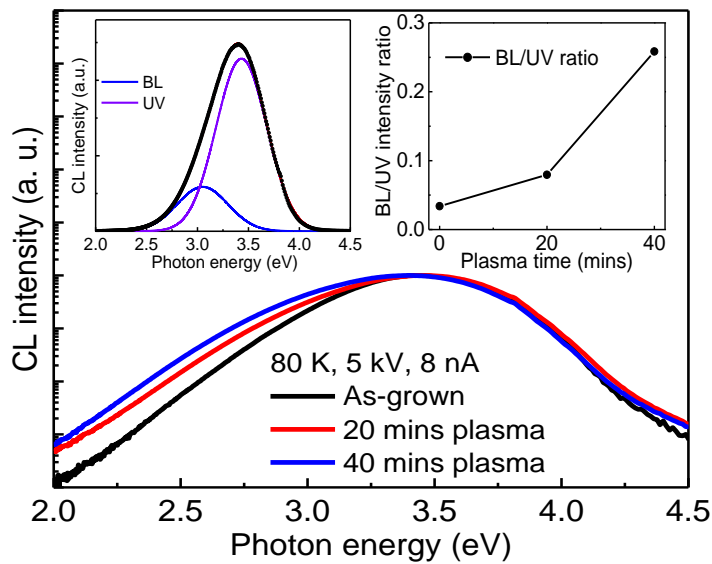

**Figure 2**. Effects of H plasma annealing on the luminescence of $Ga_2O_3$ crystal. While the UV emission and the leading edge are unaffected by the H doping, the BL is enhanced significantly. The insets show the fitted CL spectrum for the H-doped $Ga_2O_3$ crystal and intensity ratio of the UV and BL bands. The ratio increases from 0.03 for the undoped $\beta$-$Ga_2O_3$ crystal to 0.25 after 40 minutes anneal.

Temperature dependent CL was performed to determine the activation energy of the UV and BL bands. With increasing temperature from 10 K the intensity of the UV band decreases more quickly than that of the BL [Figure 3(a)], confirming they are of different chemical origin. The plot of $\ln[(I_o/I(T) - 1]$ versus $1000/T$ yields activation energies: $E_a = 48 \pm 10$ meV and $80 \pm 6$ meV for UV and BL bands, respectively [inset of Figure 3(a)]. The activation energy of the UV (48 meV) is comparable with the reported ionization energy of shallowly trapped electrons in β-$Ga_2O_3$,[3] and accordingly this value can be assigned to the binding energy of bound electrons in STEs. The activation energy of 80 meV for the BL is comparable with the ionisation energy of 110 meV for an electrically active donor in EFG β-$Ga_2O_3$ crystals,[28] but is significantly smaller than the activation energy of $V_O$ defects, being greater than 320 meV.[29] This suggests that the BL quenching is due to the thermal activation of non-radiative recombination centers. This mechanism would lead to a similar quenching trend but cannot explain why the BL and UV bands are quenched with different activation energies. Over the low temperature range (< 100 K) the both the UV and BL bands exhibit a much smaller activation energy, $E_a \approx 9$ meV. As shown in other oxide semiconductors, this small energy is likely associated with thermal quenching due to a phonon-assisted process at low temperatures.[30]

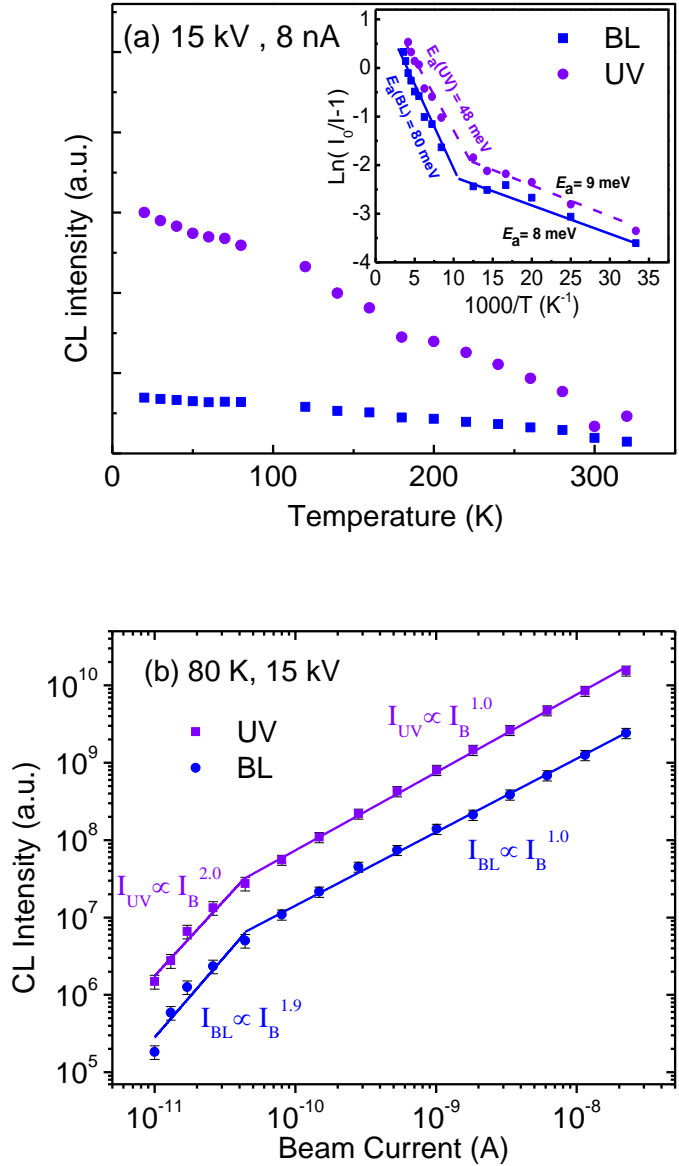

**Figure 3.** (a) Variations of UV and BL intensities with temperature for the β-Ga$_2$O$_3$ single crystal. Inset: Arrhenius analysis of the intensities yields an activation energy of 48 ± 10 and 80 ± 6 meV for the UV and BL bands, respectively. These values are associated with the ionization energy of the bound electrons in STEs and the $V_O$ defect level. (b) Dependence of UV and BL intensities as a function of CL excitation power with a primary electron beam $E_B$ = 15 keV. Power-law fits reveal that the CL intensities show a strongly super-linear dependence on beam current ($I_B$) for low excitation, with $k = 1.9 \pm 0.1$.

Next, we discuss the excitation density dependency of luminescence intensities for the CL bands. The UV and BL emissions were observed to exhibit remarkably similar excitation-power dependencies as the beam current, $I_B$ (i.e. excitation density) was increased from 10 pA to 10 nA while the beam energy was kept constant ($E_B$ = 15 keV). Varying $I_B$ in this range did not introduce any noticeable changes in peak shape or position, suggesting the CL kinetics are controlled by a competitive recombination channel, which most likely involves the capture of excited carriers to $Fe^{2+}/Fe^{3+}$ traps. The carrier capture and charge transfer processes can be described in a simplified form as:

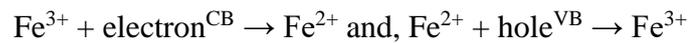

$$Fe^{3+} + electron^{CB} \rightarrow Fe^{2+} \text{ and, } Fe^{2+} + hole^{VB} \rightarrow Fe^{3+}$$

In the above equations, following the charge transfer to $Fe^{3+}$, a free hole is captured by the $Fe^{2+}$ center, which then non-radiatively relaxes into the $Fe^{3+}$ state. Although studies of deep-level traps in $Ga_2O_3$ is in its infancy, Fe is known to act as an efficient capture center in various group-II and -III compound semiconductors.[11,31] Very recently in EFG β-$Ga_2O_3$ crystals containing a similar concentration of Fe impurity, [Fe] ≈ $10^{16}$ - $10^{17}$ $cm^{-3}$, $Fe_{Ga}$ has been shown to be an energetically favourable acceptor defect and the dominant deep-level carrier trap.[10] These results provide compelling evidence for the involvement of Fe in the carrier trapping, consistent with the observed luminescence kinetics of the crystal as in the detailed analysis below. The excitation-dependent measurements of the UV and BL for the β-$Ga_2O_3$ crystal are illustrated in a log-log plot [Fig 3(b)] using a simple power-law model $I_{CL} \propto I_B^k$, where $I_{CL}$ is the CL intensity and $I_B$ is the electron beam current. Power-law fits reveal that the UV and BL intensities shows a strongly super-linear dependence on $I_B$, with $k$ = 1.9 ± 0.1 for $I_B$ < 50 pA, while the intensities exhibit a linear relationship at higher $I_B$. It is worth noting that the crystal is stable under a prolonged e-beam irradiation and the observed CL kinetics are not caused by material modification by the electron beam. The super-linear dependency at very low excitation levels in β-$Ga_2O_3$ is characteristic of dominant trapping centres being rapidly saturated by the e-beam, with the changeover at $I_B$ = 50 pA due to the complete excitation of $Fe^{3+}$ traps. Under low excitation conditions ($I_B$ < 50 pA), the carrier dynamics and CL efficiency are primarily influenced by electron trapping at $Fe^{3+}$ centres. The $Fe^{3+}$ traps are filled with electrons with increasing excitation density due to the rapid trapping of electrons by $Fe^{3+}$ and quickly saturate. As a result, additional injected carriers are then redirected to the STE and BL radiative recombination channels. This process leads to the observed super-linear increase of CL intensity in a small range of low excitation densities, hence $I_{CL} \propto I_B^{1.9}$. Once all $Fe^{3+}$ centers are completely saturated, $I_{CL}$ increases linearly with excitation power as the carrier dynamics is no longer mediated by the slow hole capturing process at $Fe^{2+}$ centers. More details on the interplay between $Fe^{3+}$ and $Fe^{2+}$ charge states are provided in the model of CL kinetics below. It has previously proposed that non-equilibrium electrons become trapped by acceptors in p-type GaN:Mg, which prevent recombination through Mg acceptor levels, leading to a gradual decrease in CL intensity with irradiation time.[32,33] For the n-type $Ga_2O_3$ crystal, the effect is opposite to the typical behaviour in p-type semiconductors because in this case hot carriers generated by CL excitation are captured by Fe ions, which simultaneously act as efficient non-radiative combination centres and inhibit carriers from participation in other recombination channels. The carrier trapping and recombination at Fe centres compete with the UV and BL recombination channels in β-$Ga_2O_3$; these processes are similar to the typical behaviour observed in many group-II and -III compound semiconductors doped with Fe.[31,34]

The observed super-linear increase of CL intensity with excitation density can be understood in terms of carrier trapping and recombination through Fe ions, which affects the CL recombination rate at dynamical equilibrium as described in the equation:[35]

$$\frac{1}{\tau_{CL}} = \frac{1}{\tau_e} + \frac{1}{\tau_h} = c_e[\text{Fe}^{3+}] + c_h[\text{Fe}^{2+}] \qquad [2]$$

Where $[Fe^{3+}]$ and $[Fe^{2+}]$ are neutral and ionised trap concentrations, respectively, with $[Fe^{3+}]$ + $[Fe^{2+}]$ = [Fe] ≈ $10^{17}$ cm$^{-3}$ for the crystal used in this work, and τ and c are the capture time (decay time) and coefficient, respectively, for electrons and holes through Fe centres. As theoretically predicted in Wickramaratne *et al.*,[34] the capture coefficient $c_e$ is about an order of magnitude higher than $c_h$. Due to the lack of detailed information about Fe traps in $\beta$-$Ga_2O_3$ and the fact that properties of Fe ions in different host materials are similar,[31,34] we will discuss the behaviour of Fe centres in analogy to those in established group-III compound semiconductors. This is justified as the $Fe^{2+}/Fe^{3+}$ charge transfer level in $\beta$-$Ga_2O_3$ is located at ~ 0.6 eV below the CBM, similar to that for Fe in GaN.[10] Prior to the CL excitation, majority of Fe ions in $\beta$-$Ga_2O_3$ are in the $Fe^{3+}$ state. $Fe^{2+}$ ions have not been detected by Electron Paramagnetic Resonance (EPR) in $\beta$-$Ga_2O_3$ crystals,[36] and are formed solely due to the ionization of $Fe^{3+}$ caused by CL excitation. It is likely that non-equilibrium carriers in CL excitation leads to the formation of an excited state of $Fe^{3+}$, similar to the electron-hole bound complex ($Fe^{3+}$, e, h) in GaN.[11] Such a shallow electron bound state is formed when the ionization of $Fe^{3+}$ does not proceed to completion as the hole in the valence band is electrostatically attracted to the Fe centre. A similar shallow bound $Fe^{3+}$ state in $Ga_2O_3$ would reduce the energy required for electrons to be emitted from $Fe^{3+}$ traps. Under low excitation conditions ($I_B < 50$ pA), the carrier dynamics and CL efficiency are primarily influenced by electron trapping to $Fe^{3+}$ centres, i.e. the first term in Eq. [2]. As the $Fe^{3+}$ traps are filled quickly with electrons with increasing excitation density due to the rapid trapping of hot electrons by $Fe^{3+}$, injected carriers are redirected to the STE and BL channels. This leads to the observed super-linear increase of CL intensity in a small range of low excitation densities ($I_B = 1 - 50$ pA).

The threshold excitation density, at which all $Fe^{3+}$ ions are at the excited state, can be estimated from the measured CL excitation conditions. The local generation rate of carriers in the interaction volume during CL measurements is approximated by:

$$\Delta n = G \frac{I_B}{e} \qquad [3]$$

In the above equation, $G$ is the e-h generation factor, $G = \frac{E_B}{E_f}(1-\eta)$, where $E_f$ is the mean energy required to create an e-h pair, $\eta$ is the electron backscattering coefficient, and $e$ is the electronic charge.[37] In general, $E_f \approx 3E_g$ for semiconductors, where $E_g$ is the band gap energy. The $\Delta n$ value calculated using Eq. [3] and $\eta = 0.26$ is ~ $2.4 \times 10^{11}$ e-h pairs/s at $I_B = 50$ pA. Using the electron interaction volume in $Ga_2O_3$ by a Monte Carlo simulation of electron trajectory,[38] we can estimate $\Delta n \approx 2 \times 10^{24}$ cm$^{-3}$.s$^{-1}$. With this excitation density most of $Fe^{3+}$ ions are transformed from the ground to excited state. The above calculation is a perfectly valid approximation when considering $\tau_e$ on the time scale of tens of ps,[39] thus $\Delta n \approx \frac{[Fe]}{\tau_e}$ under the threshold excitation conditions. For higher excitation ($\Delta n > 2 \times 10^{24}$ cm$^{-3}$.s$^{-1}$), most $Fe^{3+}$ traps are filled with electrons, thus $[Fe^{2+}] \gg [Fe^{3+}]$ and the second term dominates Eq. [2]. Capturing a hole from the valence band into $Fe^{2+}$ ions is a slow process, with $\tau_h$ on the order of μs to ms,[40] thus the hole capture rate is insignificant compared with the STE recombination rate in β-$Ga_2O_3$ (STE decay time ~ 2.1 μs).[41] Consequently, $Fe^{2+}$ does not have sufficient time to relax to its ground state, making the STE and BL the dominant recombination channels in the CL kinetics. Under this regime, we see the CL intensity increases in proportion to excitation density as shown in Figure 3(b).

## IV. CONCLUSIONS

The comprehensive cathodoluminescence analysis strongly suggests that the kinetics of charge carrier recombination and luminescence efficiency in edge-defined film-fed grown β-$Ga_2O_3$ crystals are controlled by efficient carrier trapping and competitive recombination at Fe centres, which can have an adverse effect on the performance of $Ga_2O_3$-based optoelectronic and electronic devices. The temperature-dependent properties of the UV band indicates electrons are loosely bound in self-trapped excitons with an activation energy of 48 ± 10 meV.

The behaviour of the blue emission upon the application of reducing plasma points to a donor-like defect, likely oxygen vacancies, which are strongly coupled to the crystal lattice with a Huang-Rhys factor of 7.3.

APPENDIX A: Depth-resolved CL analysis of β-Ga$_2$O$_3$ crystals

CL spectroscopy was employed to obtain depth information of luminescence signals. In this method, CL data were acquired systematically from different depths within the β-Ga$_2$O$_3$ crystal by increasing the acceleration voltage, while the beam current was adjusted so that the beam power, and hence e-h pair generation rate beam power, was kept constant. Figure 4 shows representative CL spectra collected using acceleration voltages between 2 and 20 kV, corresponding to an average sampling depth of 30 – 1400 nm as determined by the electron scattering Monte Carlo simulation.[38] The CL intensity increases rapidly with acceleration voltage while the spectral shape remains unchanged (inset), ruling out any emission related to surface effects. This result indicates both the UV and BL radiative centres are uniformly distributed throughout the crystal thickness. The reduction in the CL intensity at small sampling depths (< 300 nm) is caused by the presence of competing non-radiative defects in the near surface region.

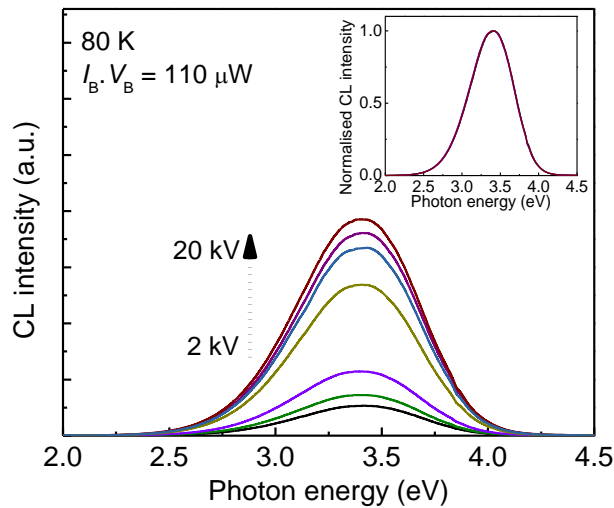

**Fig 4**. Depth-resolved CL spectra of the Ga$_2$O$_3$ crystal acquired with a constant excitation power $I_B \cdot V_B = 110$ μW at 80 K. The CL acceleration voltage was varied between 2 and 20 kV. Inset: No changes in the spectral shape or peak energy were detectable, indicating no contribution from radiative defects at the surface.

**Acknowledgements**

This work was supported under Australian Research Council (ARC) Discovery Project funding scheme (project number DP150103317).

**References**

[1] S. J. Pearton, J. C. Yang, P. H. Cary, F. Ren, J. Kim, M. J. Tadjer, and M. A. Mastro, Appl. Phys. Rev. **5**, 011301 (2018).

[2] M. Higashiwaki, K. Sasaki, H. Murakami, Y. Kumagai, A. Koukitu, A. Kuramata, T. Masui, and S. Yamakoshi, Semicond. Sci. Technol. **31**, 034001 (2016).

[3] L. Binet and D. Gourier, J. Phys. Chem. Solids **59**, 1241 (1998).

[4] Z. Hajnal, J. Miro, G. Kiss, F. Reti, P. Deak, R. C. Herndon, and J. M. Kuperberg, J. Appl. Phys. **86**, 3792 (1999).

[5] J. Varley, J. Weber, A. Janotti, and C. Van de Walle, Appl. Phys. Lett. **97**, 142106 (2010).

[6] P. Deak, Q. D. Ho, F. Seemann, B. Aradi, M. Lorke, and T. Frauenheim, Phys. Rev. B **95**, 11, 075208 (2017).

[7] J. B. Varley, H. Peelaers, A. Janotti, and C. G. Van de Walle, J. Phys.-Condes. Matter **23**, 334212 (2011).

[8] E. Chikoidze, A. Fellous, A. Perez-Tomas, G. Sauthier, T. Tchelidze, C. Ton-That, T. T. Huynh, M. Phillips, S. Russelle, M. Jennings *et al.*, Mater. Today Phys. **3**, 118 (2017).

[9] H. Gao, S. Muralidharan, N. Pronin, M. R. Karim, S. M. White, T. Asel, G. Foster, S. Krishnamoorthy, S. Rajan, L. R. Cao *et al.*, Appl. Phys. Lett. **112**, 242102 (2018).

[10] M. E. Ingebrigtsen, J. B. Varley, A. Y. Kuznetsov, B. G. Svensson, G. Alfieri, A. Mihaila, U. Badstubner, and L. Vines, Appl. Phys. Lett. **112**, 5, 042104 (2018).

[11] E. Malguth, A. Hoffmann, W. Gehlhoff, O. Gelhausen, M. R. Phillips, and X. Xu, Phys. Rev. B **74**, 12, 165202 (2006).

[12] Y. Yang, J. H. Zhang, S. B. Hu, Y. B. Wu, J. C. Zhang, W. Ren, and S. X. Cao, Phys. Chem. Chem. Phys. **19**, 28928 (2017).

[13] E. G. Víllora, K. Hatanaka, H. Odaka, T. Sugawara, T. Miura, H. Fukumura, and T. Fukuda, Solid State Commun. **127**, 385 (2003).

[14] T. Onuma, S. Fujioka, T. Yamaguchi, M. Higashiwaki, K. Sasaki, T. Masui, and T. Honda, Appl. Phys. Lett. **103**, 041910 (2013).

[15] K. Shimamura, E. G. Villora, T. Ujiie, and K. Aoki, Appl. Phys. Lett. **92**, 3, 201914 (2008).

[16] J. Varley, A. Janotti, C. Franchini, and C. G. Van de Walle, Phys.Rev.B. **85**, 081109 (2012).

[17] B. E. Kananen, N. C. Giles, L. E. Halliburton, G. K. Foundos, K. B. Chang, and K. T. Stevens, J. Appl. Phys. **122**, 6, 215703 (2017).

[18] G. C. Hu, C. X. Shan, N. Zhang, M. M. Jiang, S. P. Wang, and D. Z. Shen, Opt. Express **23**, 13554 (2015).

[19] A. Kuramata, K. Koshi, S. Watanabe, Y. Yamaoka, T. Masui, and S. Yamakoshi, Jpn. J. Appl. Phys. **55**, 6, 1202a2 (2016).

[20] C. P. Saini, A. Barman, D. Banerjee, O. Grynko, S. Prucnal, M. Gupta, D. M. Phase, A. K. Sinha, D. Kanjilal, W. Skorupa *et al.*, J. Phys. Chem. C **121**, 11448 (2017).

[21] R. T. Williams and K. S. Song, J. Phys. Chem. Solids **51**, 679 (1990).

[22] P. B. Allen and V. Perebeinos, Phys. Rev. Lett. **83**, 4828 (1999).

[23] A. Hoffmann, E. M. Malguth, and B. K. Meyer, in *Zinc oxide: from fundamental properties towards novel applications*, edited by C. F. Klingshirn *et al.* (Springer-Verlag Berlin Heidelberg 2010).

[24] T. Onuma, S. Saito, K. Sasaki, K. Goto, T. Masui, T. Yamaguchi, T. Honda, A. Kuramata, and M. Higashiwaki, Appl. Phys. Lett. **108**, 5, 101904 (2016).

[25] R. Jangir, S. Porwal, P. Tiwari, P. Mondal, S. Rai, A. Srivastava, I. Bhaumik, and T. Ganguli, AIP Adv. **6**, 035120 (2016).

[26] S. Ahn, F. Ren, E. Patrick, M. E. Law, S. J. Pearton, and A. Kuramata, Appl. Phys. Lett. **109**, 3, 242108 (2016).

[27] C. Ton-That, L. Weston, and M. R. Phillips, Phys. Rev. B **86**, 115205 (2012).

[28] A. T. Neal, S. Mou, R. Lopez, J. V. Li, D. B. Thomson, K. D. Chabak, and G. H. Jessen, Sci Rep **7**, 13218 (2017).

[29] T. C. Lovejoy, R. Y. Chen, X. Zheng, E. G. Villora, K. Shimamura, H. Yoshikawa, Y. Yamashita, S. Ueda, K. Kobayashi, S. T. Dunham *et al.*, Appl. Phys. Lett. **100**, 181602 (2012).

[30] Z. F. Lin, W. Q. Chen, R. Z. Zhan, Y. C. Chen, Z. P. Zhang, X. M. Song, J. C. She, S. Z. Deng, N. S. Xu, and J. Chen, AIP Adv. **5**, 10, 107229 (2015).

[31] E. Malguth, A. Hoffmann, and M. R. Phillips, Phys. Status Solidi B-Basic Solid State Phys. **245**, 455 (2008).

[32]    L. Chernyak, W. Burdett, M. Klimov, and A. Osinsky, Appl. Phys. Lett. **82**, 3680 (2003).

[33]    O. Lopatiuk-Tirpak, L. Chernyak, Y. Wang, F. Ren, S. Pearton, K. Gartsman, and Y. Feldman, Appl. Phys. Lett. **90**, 172111 (2007).

[34]    D. Wickramaratne, J. X. Shen, C. E. Dreyer, M. Engel, M. Marsman, G. Kresse, S. Marcinkevicius, A. Alkauskas, and C. G. Van de Walle, Appl. Phys. Lett. **109**, 4, 162107 (2016).

[35]    D. Soderstrom, S. Marcinkevicius, S. Karlsson, and S. Lourdudoss, Appl. Phys. Lett. **70**, 3374 (1997).

[36]    K. Irmscher, Z. Galazka, M. Pietsch, R. Uecker, and R. Fornari, J. Appl. Phys. **110**, 7, 063720 (2011).

[37]    B. G. Yacobi and D. B. Holt, *Cathodoluminescence microscopy of inorganic solids* (Plenum, New York, 1990).

[38]    D. Drouin, A. R. Couture, D. Joly, X. Tastet, V. Aimez, and R. Gauvin, Scanning **29**, 92 (2007).

[39]    T. Aggerstam, A. Pinos, S. Marcinkevicius, M. Linnarsson, and S. Lourdudoss, J. Electron. Mater. **36**, 1621 (2007).

[40]    R. Heitz, P. Thurian, I. Loa, L. Eckey, A. Hoffmann, I. Broser, K. Pressel, B. K. Meyer, and E. N. Mokhov, Appl. Phys. Lett. **67**, 2822 (1995).

[41]    S. Yamaoka, Y. Furukawa, and M. Nakayama, Phys. Rev. B **95**, 5, 094304 (2017).